\documentclass[aps,twocolumn,groupedaddress]{revtex4}

\usepackage{graphicx}
\usepackage{amsmath}
\usepackage{amsfonts}
\usepackage{amssymb}

\begin{document}

\def\be{\begin{equation}}
\def\ee{\end{equation}}
\def\bea{\begin{eqnarray}}
\def\eea{\end{eqnarray}}
\def\bma{\begin{mathletters}}
\def\ema{\end{mathletters}}
\newcommand{\one}{\mbox{$1 \hspace{-1.0mm}  {\bf l}$}}
\newcommand{\eins}{\mbox{$1 \hspace{-1.0mm}  {\bf l}$}}
\def\C{\hbox{$\mit I$\kern-.7em$\mit C$}}
\newcommand{\tr}{{\rm tr}}
\newcommand{\shalf}{\mbox{$\textstyle \frac{1}{\sqrt{2}}$}}
\newcommand{\ket}[1]{ | \, #1  \rangle}
\newcommand{\bra}[1]{ \langle #1 \,  |}
\newcommand{\proj}[1]{\ket{#1}\bra{#1}}
\newcommand{\kb}[2]{\ket{#1}\bra{#2}}
\newcommand{\bk}[2]{\langle \, #1 | \, #2 \rangle}
\def\II{I(\{p_k\},\{\rho_k\})}
\def\ss{{\cal K}}
\tolerance = 10000

\bibliographystyle{apsrev}

\title{$\frac{1}{2}$--Anyons in small atomic Bose-Einstein condensates}

\author{B. Paredes}\email[]{Belen.Paredes@uibk.ac.at} \homepage[]{www.uibk.ac.at}
\author{P. Fedichev}
\author{J. I. Cirac}
\author{P. Zoller}

\affiliation{Institute for Theoretical Physics, University of
Innsbruck, Austria}

\begin{abstract}
We discuss a way of creating, manipulating and detecting anyons in
rotating Bose-Einstein condensates consisting of a small number of
atoms . By achieving a quasidegeneracy in the atomic motional
states we drive the system into a $\frac{1}{2}$--Laughlin state
for fractional quantum Hall bosons. Localized
$\frac{1}{2}$--quasiholes can be created by focusing lasers at the
desired positions. We show how to manipulate these quasiholes in
order to probe directly their $\frac{1}{2}$--statistics.
\end{abstract}

\date{\today}
\pacs{PACS} \maketitle


The experimental achievement of Bose--Einstein condensation of
weakly interacting atomic gases \cite{BEC} promises new
possibilities to study the quantum properties of many--body
systems. As compared to other systems, quantum degenerate atomic
gases can be easily controlled and manipulated by electromagnetic
fields, which makes them ideal candidates for the study of several
intrinsically quantum phenomena. Until now, interest has focused
mainly on using atomic condensates to study single particle
quantum phenomena (as those occurring in atomic interferometry or
atom optics), since for low temperatures atoms in condensates
occupy essentially the same single particle state. In contrast,
the possibility of observing entanglement, one of the most
fascinating features of quantum mechanics, remains almost
unexplored, requiring a way of making the gas to {\em effectively}
behave in a {\em strongly interacting} manner. First steps in this
direction are the recent proposals to entangle atomic beams and
atomic spin squeezing in Bose-Einstein condensates (BEC)
\cite{entangle}, and  the numerical prediction of new correlated
phases appearing in rotating condensates \cite{Wilkin}.

In this Letter we show how a quantum degenerate Bose gas
consisting of a small number of atoms can be used to study a
quantum phenomenon highly collective in nature, namely the
formation of quasiparticles exhibiting fractional statistics. This
system offers the novel possibility of creating and manipulating
anyons in a well controlled way that may allow for the
experimental probing of their fractional statistics. The idea is
to achieve a quasidegeneracy of the atomic motional states by
rotating the trap that confines the atoms, as it has successfully
been done for the creation of vortices \cite{Vortices}. Under
appropriate conditions (like two dimensional confinement), the
situation we describe can be understood in terms of the fractional
quantum Hall effect \cite{QHE} for bosons \cite{DasSarma}. As in
the theory of the fractional quantum Hall  for electrons
\cite{Mac1}, elementary excitations exhibiting fractional
statistics appear. We consider a situation where the atomic system
is first prepared in a $\frac{1}{2}$--Laughlin state, a highly
correlated quantum liquid with nearly uniform density. We then
show that piericing the system by offresonant laser light a single
Laughlin quasihole localized at some chosen position $z_{0}$ can
be created. This excitation involves a density profile in which
exactly $\frac{1}{2}$--atom has been removed at $z_{0}$. In
addition, lasers provide us with a tool for creating states with
two quasiholes at the desired positions, and moreover, for
creating superpositions of states having one and two quasiholes.
Driving the superposition state along the proper path, as in a
Ramsey-type interferometer, will allow to test the fractional
statistical phase. Apart from interest of detecting such a phase
directly, the ideas developed in the present work may pave the way
for a physical implementation for quantum information processing
based on anyons; as reported in \cite{fault}, the most robust way
of performing quantum computations seems to be based on
excitations with fractional statistics since they have several
fault--tolerant properties built--in. On the other hand, the
conditions required to observe the effects predicted here are very
demanding, and we expect that can be reached in the near future
only with a small number of atoms.

We consider a set of $N$ bosonic atoms confined in a potential
which rotates in the $x-y$ plane at a frequency $\Omega$. We will
assume that the confinement in the $z$ direction is sufficiently
strong so that we can ignore the excitations in that direction,
and we will consider that the potential in two dimensions is
isotropic and harmonic. The Hamiltonian describing this situation
in a frame rotating with the trap is: 
\begin{equation} 
H=\frac{1}{2}
\sum _{i=1}^{N} \left( - \nabla _{i}^2 + r_{i}^2  -
2\frac{\Omega}{\omega}L_{iz}\right) + \eta \sum_{i<j}^{N}\delta
({\bf r_{i}} - {\bf r_{j}}), 
\label{H2D} 
\end{equation} 
with $L_{iz}$ being
the $z$ component of the angular momentum of the $i$--th atom, and
where we have used the trap energy, $\hbar \omega$, as the unit of
energy, and  $\ell=( \hbar/ m \omega  )^{1/2}$ as the unit of
length. The atoms are interacting via an effective contact
potential, and the interacting coupling constant $\eta$ is related
to the $s$--wave scattering length, $a$, and to the localization
length in the $z$ direction, $\ell_{z}$, by $\eta=\sqrt{2/ \pi}a /
\ell_{z}$.

As pointed out recently \cite{Wilkin}, in the limit $\Omega=
\omega$ the Hamiltonian (\ref{H2D}) is formally identical to the
Hamiltonian of electrons in the quantum Hall effect \cite{QHE}
with $\omega$ playing the role of the magnetic field (with
cyclotron frequency $\omega_{c}= 2\omega$), and the usual Coulomb
interaction between electrons replaced by a contact interaction.
Regardless that we are dealing with bosons instead of fermions the
two problems are formally identical in this limit. We will now
study how to create and manipulate anyons \cite{Mac1} in our twin
atomic system.

We begin by writing the Hamiltonian (\ref{H2D}) in the following
form: 
\begin{equation} 
H=H_{B}+H_{L}+ V. 
\end{equation} 
Here, $H_{B}=\sum_{i=1}^{N} -
\nabla_{i}^2/2 +r_{i}^2/2 - L_{iz}$ is the quantum Hall single
particle Hamiltonian, whose single particle energy levels are the
Landau levels equally spaced by the cyclotron energy
$2\hbar\omega$. The Hamiltonian $H_{L}=(1-\Omega/\omega)L_{z}$  is
proportional to the $z$ component of the total angular momentum,
$L_{z}=\sum_{i=1}^NL_{iz}$, and $V$ is the interaction term. From
now on we consider the limit in which the energy scales
characterizing Hamiltonians $H_{B}$ and $V$ are much larger than
the one corresponding to $H_{L}$. This means that both the trap
energy and the typical interaction energy are large compared to
the angular momentum term. In this limit the ground state and
elementary excitations of the system will lie on the subspace of
common  zero energy eigenstates of  $H_{B}$ and $V$. We now derive
the spatial form of the many-body wave functions  $\Psi$ lying
within  this subspace. In order to be a zero energy eigenstate of
$H_{B}$, $\Psi$ must lie within the subspace generated by tensor
products of lowest Landau level single particle states
\cite{Mac1}, 
\begin{equation} 
\Psi[z] =P(z_{1}, \ldots , z_{N})
\prod_{k}e^{-\vert z_{k} \vert^{2}/4}, 
\label{form1} 
\end{equation} 
where
$P[z]$ is a polynomial in each of the atomic coordinates
$z_{k}=x_{k}+iy_{k}$. Let's assume that $\Psi[z]$ is also an
eigenstate of $V$ with eigenvalue zero, and let's choose any pair
of particles $i$ and $j$. The dependence of $\Psi[z]$ on $z_{i}$
and $z_{j}$ can be reexpressed in terms of the relative and center
mass coordinates, $z_{ij}$, $Z_{ij}$, so that we can expand
$P[z]=\sum_{m} z_{ij}^{m}F_{m}$, where $F_{m}$ depends on $Z_{ij}$
and on the positions of all the other particles. As we are dealing
with bosons only even values of $m$ appear in the sum. In order
for $\Psi[z]$ to be annihilated by the hard-core interaction $V$,
$F_{0}$ must be identically zero. It follows that $z_{ij}^{2}$ is
a factor of $P[z]$ so that 
\begin{equation} 
P[z] = Q[z] \prod _{i<j}
(z_{i}-z_{j})^{2}\;. 
\label{form2} 
\end{equation}

We then diagonalize Hamiltonian $H_{L}$ within the truncated
Hilbert space of wave functions of the form specified by
(\ref{form1}) and (\ref{form2}). We note that when $P[z]$ is a
homogeneous polynomial in $[z]$ the state (\ref{form1}) is an
eigenstate of $H_{L}$ with eigenvalue $E(M)=(1-\Omega/ \omega)M$,
where the total angular momentum $M$ equals the homogeneous degree
of $P[z]$. It follows that the ground state of the system is the
state with the lowest angular momentum, that is, the one with
$Q[z]=1$: 
\begin{equation} 
\psi[z] = \prod _{i<j} (z_{i}-z_{j})^{2}
\prod_{k}e^{-\vert z_{k} \vert^{2}/4}.
\label{Lau} 
\end{equation} 
This state is the bosonic variant of the
Laughlin wave function for quantum Hall electrons \cite{Lau}.

\begin{figure}
\begin{center}
\includegraphics[height=4cm,width=5cm]{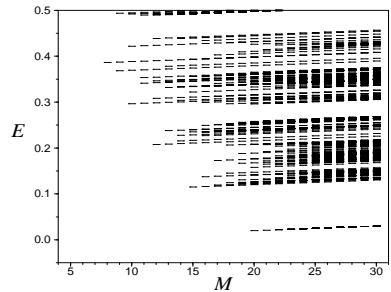}
\end{center}

\caption{Eigenvalues of the Hamiltonian  for  $N=5$ bosons as a
function of the total angular momentum $M$. For illustrating
purposes we chose $(1-\Omega/\omega)/\eta=0.001$. Energy is
measured in units of $\eta \hbar \omega $.} \label{fig1}

\end{figure}

We have confirmed the above arguments by diagonalizing the
Hamiltonian (\ref{H2D}) exactly for $N=5$ bosons. Fig.~
(\ref{fig1}) shows the energy spectrum. There is a branch of
states well separated in energy from the rest of the spectrum.
These states are polynomial states of the form given by
(\ref{form1}) and (\ref{form2}), and the ground state is the five
atoms Laughlin state (\ref{Lau}), with angular momentum
$M_{0}=N(N-1)/2=20$.

\begin{figure}

\begin{center}

\includegraphics[height=6cm,width=6cm]{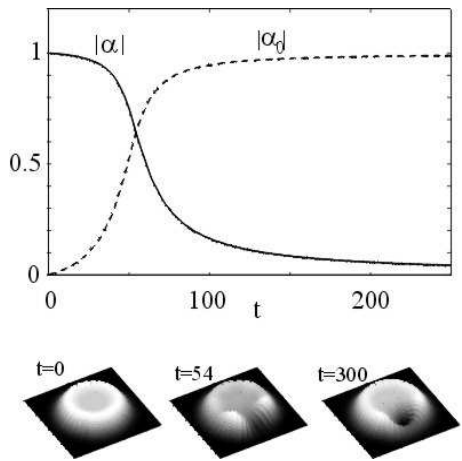}

\end{center}

\caption{ Coefficients of the ground state for the system of $N=5$
particles as a function of time, in the Laughlin state $\psi$
(filled line), and in the quasihole state $\psi_{z_{0}}$ (dashed
line). The laser is localized at $z_{0}=2.5 \ell$ (the size of the
droplet is $ \sim 4.2\ell $), and its  intensity increases with time
as $V_{0}=0.1t$ (time in units of $\omega^{-1}$). The three plots
at the bottom show the density profile of the ground state at
three steps of the time evolution (the size of the plane is $6\ell
\times 6\ell$).} \label{fig2}
\end{figure}

To create a single fractional quasiparticles in the Laughlin state
we insert a laser localized (within an area $\sim \ell^2$) at some
position $z_{0}$. We require $\vert z_{0}\vert$ to be within the
size of the Laughlin state (~ $\sim 2\sqrt{N-1}$). The
presence of the laser can be described by a localized repulsive
potential, so that the new Hamiltonian of the system can be
approximated by: 
\begin{equation} 
H_{\circ}=H_{L}+V_{0}\
\sum_{i}\delta(z_{i}-z_{0}). 
\label{Hdel1} 
\end{equation} 
We have studied the
time evolution of the ground state as the intensity $V_{0}$ of the
laser increases with time, under the assumption that the system
remains always in the truncated Hilbert space of polynomial wave
functions of the form (\ref{form2}). We note that the total
angular momentum $L$ does not commute anymore with the Hamiltonian
(\ref{Hdel1}), since the $\delta$-potential breaks the rotational
symmetry. We have solved the dynamics exactly for the model system
of $N=5$ bosons. Fig.~\ref{fig2} shows the time evolution of the
ground state. For low intensities of the laser the angular
momentum term dominates and the system remains in the Laughlin
state. But when the laser power becomes sufficiently large the
system evolves to a Laughlin quasihole state with wave function
\begin{equation} 
\psi_{z_{0}}[z]=\prod_{i}(z_{i}-z_{0}) \ \psi[z], 
\end{equation} 
where
$\psi[z]$ is the Laughlin wave function (\ref{Lau}). Note that the
state $\psi_{z_{0}}$ is a superposition of homogeneous polynomial
states with angular momenta running from $M_{0}$ to $M_{0}+N$. In
this quasihole state all the particles are expelled from the
position where the laser is located, so that the potential energy
is minimized. This gain in potential energy compensates the cost
of confining energy due to the spreading of the gas.

It is remarkable that for intermediate laser intensities the
ground state is approximately equal to a superposition of the
Laughlin state and the quasihole state: $\Psi \sim \alpha \psi +
\alpha_{0} \psi_{z_{0}}$, with no other states participating
significantly in the evolution of the system. To understand this
behavior let us consider a state quasi-orthogonal to this
subspace, that is a state having a quasihole localized at some
other position $z_{\ell} \ne z_{0}$. To reach this state the
system has to pay confining energy (because of the system
spreading around $z_{\ell}$), but there is no gain in potential
energy since the hole has been created at the wrong position. It
follows that any state out of the subspace generated by $\psi$ and
$\psi_{0}$ will be always much higher in energy (no matter the
intensity of the laser), and will not take part in the evolution
of the system.

Based on the creation of quasiholes and superposition states we
describe now a possible experiment to reveal directly the
statistics of anyons in a Ramsey-type interferometer:

1){\em First step: Preparation of the initial state.} We focus a
laser at position $z_{0}$ and increase its intensity until a
single quasihole is created. Keeping constant the intensity of
this laser we then adiabatically insert another laser at position
$z_{1}$, far enough from $z_{0}$ ($\vert z_{1} - z_{0} \vert \geq
\ell $). The new Hamiltonian is: $H_{\circ \circ}=H_{L}+V_{0}\
\sum_{i}\delta(z_{i}-z_{0}) + V_{1}\ \sum_{i}\delta(z_{i}-z_{1})$.
Following an analogous pattern to the one shown in Fig.~
\ref{fig2} for a single quasihole, the system now evolves from the
one-quasihole state to a two-quasihole state of the form:
$\psi_{z_{0},z_{1}}[z]=\prod_{i}(z_{i}-z_{0}) (z_{i}-z_{1}) \
\psi[z]$.  The crucial point is that at a certain point of the
evolution the system reaches a superposition state $\Psi \sim
(\psi_{z_{0}}+\psi_{z_{0},z_{1}})$. This is precisely the
superposition we need to test the statistical angle. We then stop
the evolution of the system at this point, and remove
instantaneously the laser located at $z_{1}$.  In this way the
state of the system remains unchanged but we go back to
Hamiltonian $H_{\circ}$. Note that the state $\Psi$ is no longer
an eigenstate of this Hamiltonian. We are now in a position to
detect the statistical phase.

2){\em Second step: Statistical phase accumulation.} We
adiabatically move the remaining laser along a path enclosing
position $z_{1}$, so that the time dependent Hamiltonian is:
$H(t)=H_{L}+V_{0}\sum_{i}\delta(z_{i}-z(t))$,  with
$z(t)-z_{1}=\vert z_{0}-z_{1}\vert e^{i\beta t}$. As follows from
the adiabatic theorem \cite{Berry} the evolved state at  the end
of the process will be: 
\begin{equation}
\Psi^{\prime}=e^{-iE_{1}T+\varphi_{1}}\psi_{z_{0}}+
e^{-iE_{2}T+\varphi_{2}}\psi_{z_{0},z_{1}}, 
\label{Hrot} 
\end{equation} 
where
besides the dynamical phase each state picks up a Berry phase. The
difference between the Berry phases $\varphi= \varphi_{1}-
\varphi_{2}$ of the two states reflects the extra phase that the
quasihole at $z_{0}$ picks up due to the presence of the other
quasihole at $z_{1}$, thus it is the statistical phase.

We have simulated this experiment with our $5$ bosons model
system. In order to have a more precise detection of the Berry
phase we have considered a regime in which the dynamical relative
phase is much smaller than the relative Berry phase, so that
$(E_{1}-E_{2})/\beta \sim 5(1-\omega/\Omega)\ll 1$. We have
performed the simulation for different closed paths and different
positions of the two quasiholes. The relative phase emerging from
the numerical calculation is $\varphi=1.031\pi$, so that
$\Psi^{\prime} \sim \psi_{z_{0}}-\psi_{z_{0},z_{1}}$. Since the
closed loop we have performed is equivalent to  two consecutive
interchanges, this result states that, as it should be,
$\frac{1}{2}$-- quasiholes pick up a phase
$\varphi=\frac{1}{2}\pi$ when we interchange them. Thus, they do
not behave as bosons nor fermions, but as anyons with statistics
$\frac{1}{2}$.

\begin{figure}

\includegraphics[height=5cm,width=7cm]{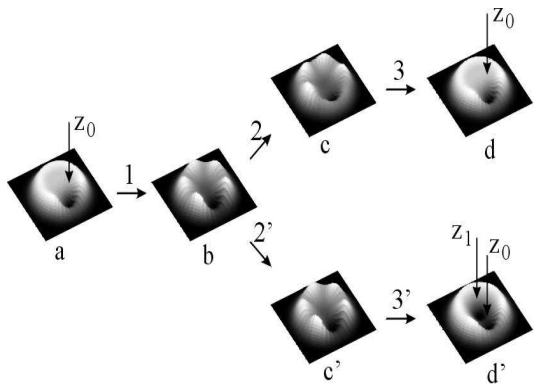}

\caption{Numerical simulation of the Ramsey--like experiment for
$N=5$ bosons. {\em Step 1}: Adiabatically inserting a laser at
position $z_{1}$ we drive the system from $\psi_{z_{0}}$ (plot a)
into $\Psi=\psi_{z_{0}}+ \psi_{z_{0},z_{1}}$ (plot b). {\em Step
2}: We instantaneously remove the laser at $z_{1}$ and drive the
laser at $z_{0}$ along a closed path enclosing $z_{1}$. The
resulting state is $\Psi'\sim \psi_{z_{0}}- \psi_{z_{0},z_{1}}$
(plot c). {\em Step 3}: We restore the laser at $z_{1}$ and
continue increasing its intensity. The final state (plot d) is the
single quasihole state $\psi_{z_{0}}$. In contrast, if the laser
at $z_{0}$ makes two complete loops around $z_{1}$ (step 2'). In
this case the resulting state is again $\Psi\sim \psi_{z_{0}}+
\psi_{z_{0},z_{1}}$ (plot c'), which evolves to the state with two
holes (plot d').} \label{fig3}

\end{figure}

3){\em Third step: Detecting the statistical phase.} Suppose that
we come back to the point at which we stopped the  evolution of
the system. This means that we instantaneously restore the laser
we had removed at position $z_{1}$ and continue increasing its
intensity. If we had not made the {\em step 2} of the experiment,
the system would be in the state $\Psi$ and would evolve to the
two-quasihole state $\psi_{z_{0},z_{1}}$. However, after
performing {\em step 2} the state $\Psi$ has changed into
$\Psi^{\prime}$ and the system will now evolve in a different way.
Fig.~ \ref{fig3} shows how the minus sign is reflected in the
density profile of the final state. Instead of getting a two hole
state we go back to the state with only one quasihole at $z_{0}$.

We discuss now the  set of conditions that a system of $N$ atoms
must fulfill to perform an experiment as the one we have described
above. First of all, we have made a two-dimension approximation,
so that we need $\ell \ll \ell_{z}$. We have considered also a
contact interaction between the atoms and this requires $a <
\ell_{z}$ \cite{Petrov}. We have created a state with two Laughlin
quasiholes. This implies a total angular momentum $M=M_{0}+2N$ and
thus an angular momentum energy per particle $e_{L}\lesssim
2(1-\Omega/\omega)N$. Remember that we have projected onto the
subspace of common zero eigenstates of $H_{B}$ and $V$. In order
for this projection to be valid  we then need $e_{L}$ to be much
smaller than both the trap energy and the typical interaction
energy, so that the conditions $2(1-\Omega/\omega)N \ll 1, \
\eta /4\pi $ are required \cite{Belen2}. For creating the
$\frac{1}{2}$--quasiholes we had to focus the lasers within a
distance $\sim \ell$. For a localization length $\ell \sim 1\mu m$
this implies an upper limit for the trap frequency of $\sim 1kHz$.
The preparation of  the superposition state used to test the
statistical angle requires to adiabatically increase the laser
intensity. In order to estimate how slowly the laser needs to be
inserted for a system of $N$ particles we have used a two state
approximation. Confining ourselves to a basis formed by states
$\psi_{z_{0}}$ and $\psi_{z_{0},z_{1}}$, we find that the width of
the avoided crossing is $ \Delta\sim N (1-\Omega/\omega)(\vert
z_{1} \vert/\sqrt{N})^N. $ This is normally a small quantity,
quickly decreasing when the hole is made closer to the center of
the trap. The time scale $\Delta^{-1}$ gives us the estimation on
how slowly the laser is to be set on/off in order to reach the
necessary adiabaticity requirements. Finally, the most restrictive
condition is the temperature. In order to freeze out the
excitations we need $kT/\hbar\omega \ll (1-\Omega/\omega)$,
which together with the above conditions
implies $kT/\hbar\omega \ll 1/N, \eta/N$
\cite{ions}.

In conclusion, this letter provides a way of creating anyons in
rotating Bose Einstein condensates, and proposes an experiment in
which the fractional statistics can be tested. While the
requirements are very demanding, we expect that creation and
(effectively) ground state cooling of small ensembles of bosonic
atoms to be within experimental reach in coming years. Development
of these experimental techniques promises the controlled
engineering of strongly correlated, entangled states of atoms,
with novel applications in quantum information.

We thank L. Pitaevskii, G. Shlyapnikov and C. Tejedor for
discussions. Work supported by the Austrian Science Foundation,
EQUIP, ESF, the European TMR networks, and the Institute for
Quantum Information Gmbh.

\end{document}